\documentclass[aps,twocolumn,pra,showpacs]{revtex4}
\usepackage{graphicx}
\usepackage{amsmath,amssymb,amsthm,bbm,latexsym}
\usepackage{amsthm}
\usepackage{amsbsy}
\usepackage{epsfig}
\usepackage{graphicx}
\usepackage{float}
\usepackage{dcolumn}
\usepackage{amsmath}

\begin{document}

\author{Yong-Cheng Ou and Heng Fan}

\affiliation{Institute of Physics, Chinese Academy of Sciences,
Beijing 100080, People's Republic of China}

\title{Monogamy Inequality in terms of Negativity for Three-Qubit States}

\begin{abstract}
We propose a new entanglement measure to quantify three qubits
entanglement in terms of negativity. A monogamy inequality analogous
to Coffman-Kundu-Wootters (CKW) inequality is established. This
consequently leads to a definition of residual entanglement, which
is referred to as three-$\pi$ in order to distinguish from
three-tangle. The three-$\pi$ is proved to be a natural entanglement
measure. By contrast to the three-tangle, it is shown that the
three-$\pi$ always gives greater than zero values for pure states
belonging to the $W$ and GHZ classes, implying there always exists
three-way entanglement for them, and three-tangle generally
underestimates three-way entanglement of a given system. This
investigation will offer an alternative tool to understand genuine
multipartite entanglement.
\end{abstract}


\maketitle

\section{Introduction}

Quantum entanglement lies at the heart of quantum information
processing and quantum computation\cite{nielsen}, accordingly its
quantification has drawn much attention in the last decade. In order
for such quantification legitimate measures of entanglement are
needed as a first step. The existing well-known bipartite measure of
entanglement with an elegant formula is the concurrence derived
analytically by Wootters\cite{wootters} and the entanglement of formation%
\cite{bennett, hill} is a monotonically increasing function of
concurrence. They have been applied to describing quantum phase
transition in various interacting quantum many-body
systems\cite{osterloh,wu}. Another useful entanglement measure is
negativity\cite{we}, regarded as a quantitative version of Peres'
criterion for separability. Comparing with the concurrence, the
process calculating the negativity is significantly simplified with
respect to mixed states since it does not need the convex-proof
extension.

On the other hand, multipartite entanglement is a valuable physical
resource in large-scale quantum information processing\cite{bri} and
plays an important role in condensed matter physics. The negativity
is used to study multipartite entanglement in a Fermi
gas\cite{lunkes}. However, it is a formidable task to quantify
multipartite entanglement since there is few well-defined
multipartite entanglement measures just like for bipartite systems.
As for now, the widely-used basis for characterizing and quantifying
tripartite entanglement is the three-tangle \cite{coffman}. Very
recently the proof for the general CKW inequality for bipartite
entanglement\cite{osborne} and the demonstration that the CKW
inequality cannot generalize to higher-dimensional systems\cite{ou1}
have been provided.

Recall that the concurrence of a two-qubit state $\rho$ is defined
as $\mathcal{C}(\rho)\equiv \max\{0,
\sqrt{\lambda_1}-\sqrt{\lambda_2}-\sqrt{\lambda_3}-\sqrt{\lambda_4}\}$,
in which $\lambda_1,...,\lambda_4$ are the eigenvalues of the matrix
$\rho(\sigma_y\otimes\sigma_y)\rho^*(\sigma_y\otimes\sigma_y)$ in
nonincreasing order and $\sigma_y$ is a Pauli spin matrix.  For a
pure three-qubit state $\rho_{ABC}$, the CKW inequality in terms of
concurrence can read
\begin{equation}
\mathcal{C}_{AB}^{2}+\mathcal{C}_{AC}^{2}\leq
\mathcal{C}_{A(BC)}^{2}, \label{a}
\end{equation}%
where $\mathcal{C}_{AB}$ and $\mathcal{C}_{AC}$ are the concurrences
of the mixed states
$\rho_{AB}=\mathrm{Tr}_{C}(|\phi\rangle_{ABC}\langle\phi|)$ and $\rho_{AC}=\mathrm{Tr}_{B}(|\phi\rangle_{ABC}%
\langle\phi|)$, respectively, and
$\mathcal{C}_{A(BC)}=2\sqrt{\det\rho_A}$ with
$\rho_{A}=\mathrm{Tr}_{BC}(|\phi\rangle_{ABC}\langle\phi|)$.
According to Eq.(\ref{a}) the three-tangle can be defined as
\begin{equation}\label{22}
\tau_{ABC}=C_{A(BC)}^{2}-C_{AB}^{2}-C_{AC}^{2},
\end{equation}
which is used to characterize three-way entanglement of the
state\cite{due}. For example, quantified by the three-tangle, the
state $|\mathrm{GHZ}\rangle =\frac{1}{\sqrt{2}}(|000\rangle
+|111\rangle )$ has only three-way entanglement, while the state
$|W\rangle =\frac{1}{\sqrt{3}}(|100\rangle +|010\rangle +|001\rangle
)$ has only two-way entanglement. For a general mixed three-qubit
state of $\rho_{ABC}$ the three-tangle should be
$\tau_{ABC}=\min\left[C_{A(BC)}^{2}\right]-C_{AB}^{2}-C_{AC}^{2}$,
where $C_{A(BC)}^{2}$ has to be minimized for all possible
decomposition of $\rho_{ABC}$.
Now one may wonder whether there exist other entanglement measures satisfying Eq.(\ref{a}%
) and whether the three-way entanglement of a given state provided by these entanglement measures is the same. This will help us further understand genuine multipartite entanglement.

To this end, the main result of this paper is to provide a monogamy
inequality in terms of negativity. In Sec II, we recall some basic
concepts of the negativity. In Sec III, we deduce the monogamy
inequality in terms of negativity. In Sec IV and V, the three-$\pi$
analogous to three-tangle is defined, which is shown to be a
natural entanglement measure. By calculation on the $|W\rangle$ state, the $|%
\mathrm{GHZ}\rangle$ state, and the superposed states of the two states, the three-$%
\pi$ is shown to be greater than zero, i.e., for such states there
always exists three-way entanglement. It is also shown that the
three-$\pi$ is always not less than the three-tangle for any
tripartite pure states and can be extended to mixed three-qubit
states and general pure $n$-qubit states. The conclusions are in Sec
VI.

\section{Basic concepts of the negativity }For a either pure or mixed state $\rho $ in the tensor
product $\mathcal{H}_{A}\otimes \mathcal{H}_{B}$ of two Hilbert
spaces $\mathcal{H}_{A}$, $\mathcal{H}_{B}$ for two subsystems $A$
and $B$, the partial transpose with respect to $A$ subsystem is
$(\rho
^{T_{A}})_{ij,kl}=(\rho )_{kj,il}$ and the negativity is defined by $%
\mathcal{N}={(\Vert \rho ^{T_{A}}\Vert -1})/{2}$ where the trace norm $\Vert
R\Vert $ is given by $\Vert R\Vert =\mathtt{Tr}\sqrt{RR^{\dagger }}$. $%
\mathcal{N}>0$ is the necessary and sufficient inseparable condition for the
$2\otimes 2$ and $2\otimes 3$ bipartite quantum systems\cite{ho}. In order
for any maximally entangled state in $2\otimes 2$ systems to have the
negativity one, it can be reexpressed as
\begin{equation}
\mathcal{N}=\Vert \rho ^{T_{A}}\Vert -1,  \label{c}
\end{equation}%
with only a change of the constant factor 2. Therefore $%
\mathcal{N}=1$ for Bell states like $\frac{1}{\sqrt{2}}(|01\rangle
+|01\rangle )$ and vanishes for factorized states. For pure two-qubit
systems in terms of the coefficients $\{\phi _{00},\phi _{01},\phi
_{10},\phi _{11}\}$ of $|\phi _{AB}\rangle $ with respect to an orthonormal
basis the concurrence is defined as $\mathcal{C}_{AB}=2|\phi _{00}\phi
_{11}-\phi _{01}\phi _{10}|$. From Eq.(\ref{c}) it is easy to check that $%
\mathcal{N}_{AB}=\mathcal{C}_{AB} $ for such systems. Now let us consider
pure three-qubit systems $A$, $B$, and $C$ in the standard basis $%
\{|ijk\rangle \}$, where each index takes the values 0 and 1: $|\phi \rangle
_{ABC}=\sum_{ijk}\phi _{ijk}|ijk\rangle $. For our goal it is necessary to
show $\mathcal{N}_{A(BC)}=\mathcal{C}_{A(BC)}$. The density matrix of $|\phi \rangle _{ABC}$ is $\rho =|\phi \rangle
_{ABC}\langle \phi |$ and $\rho ^{T_{A}}=\sum_{ijk,i^{\prime }j^{\prime
}k^{\prime }}\phi _{ijk}\phi _{i^{\prime }j^{\prime }k^{\prime }}^{\ast
}|i^{\prime }jk\rangle \langle ij^{\prime }k^{\prime }|$. Following from Eq.(%
\ref{c}) we arrive at
\begin{eqnarray}
\mathcal{N}_{A(BC)} &=&\Vert \sum_{ijk,i^{\prime }j^{\prime }k^{\prime
}}\phi _{ijk}\phi _{i^{\prime }j^{\prime }k^{\prime }}^{\ast }|i^{\prime
}jk\rangle \langle ij^{\prime }k^{\prime }|\Vert -1  \notag \\
&=&\Vert \sum_{ijk}\phi _{ijk}|jk\rangle \langle i|\otimes \sum_{i^{\prime
}j^{\prime }k^{\prime }}\phi _{i^{\prime }j^{\prime }k^{\prime }}^{\ast
}|i^{\prime }\rangle \langle j^{\prime }k^{\prime }|\Vert -1  \notag \\
&=&\Vert R\otimes R^{\dag }\Vert -1 =\Vert R\Vert ^{2}-1 \notag \\
&=&2\sqrt{\lambda _{0}\lambda _{1}}=\mathcal{C}_{A(BC)},  \label{g}
\end{eqnarray}%
where $R=\sum_{i^{\prime }j^{\prime }k^{\prime }}\phi _{i^{\prime }j^{\prime
}k^{\prime }}^{\ast }|i^{\prime }\rangle \langle j^{\prime }k^{\prime }|$, $%
\lambda _{0}$ and $\lambda _{1}$ are eigenvalues of $RR^{\dag}$. The
obtaining of the third formula is based on the property of the trace norm $%
\Vert G\otimes Q\Vert =\Vert G\|\cdot \Vert Q\Vert $,
observation that $RR^{\dag }=\sum_{i^{\prime }j^{\prime
}k^{\prime },ijk}\phi _{ijk}\phi _{i^{\prime }j^{\prime }k^{\prime }}^{\ast
}|i^{\prime }\rangle \langle j^{\prime }k^{\prime }|\cdot |jk\rangle \langle
i|$, and $\|R\|$ is equal to the sum of the square root of eigenvalues $%
\lambda_{i}$ of $RR^{\dag}$ with $\lambda_{0}+\lambda_{1}=1$.
From another observation that $\lambda_0$ and $\lambda_{1}$ are also the
eigenvalues of the reduced density matrix $\rho _{A}=\mathrm{Tr}_{BC}(|\phi
\rangle _{ABC}\langle \phi|)$ whose matrix elements are $\mu_{00}=\sum_{jk}%
\phi _{0jk}\phi _{0jk}^{\ast}, \mu_{01}=\sum_{jk}\phi _{0jk}\phi
_{1jk}^{\ast }, \mu_{10}=\sum_{jk}\phi _{1jk}\phi _{0jk}^{\ast }$ and $%
\mu_{11}=\sum_{jk}\phi _{1jk}\phi _{1jk}^{\ast }$, and the concurrence
between $A$ and $BC$ is defined as $\mathcal{C}_{A(BC)}=\sqrt{2(1-\mathrm{Tr}%
\rho_A^2)}=2\sqrt{\lambda_0\lambda_1}$, the last formula is
obtained. The next paragraphs are devoted to one of the main results
of this paper.

\section{Monogamy inequality in terms of negativity}
For any pure
$2\otimes 2 \otimes 2$ states $|\phi\rangle_{ABC}$, the entanglement
quantified by the negativity between $A$ and $B$, between $A$ and
$C$, and between $A$ and the single object $BC$ satisfies the
following CKW- inequality-like monogamy inequality
\begin{equation}  \label{g1}
\mathcal{N}^2_{AB}+\mathcal{N}_{AC}^2\leq \mathcal{N}^2_{A(BC)},
\end{equation}
where $\mathcal{N}_{AB}$ and $\mathcal{N}_{AC}$ are the negativities
of the mixed states
$\rho_{AB}=\mathrm{Tr}_{C}(|\phi\rangle_{ABC}\langle\phi|)$ and $\rho_{AC}=\mathrm{Tr}_{B}(|\phi\rangle_{ABC}%
\langle\phi|)$, respectively.

In order to prove Eq.(\ref{g1}) it is helpful to recall the Theorem
appearing in\cite{chen}, which states that for any $m\otimes n(m\leq n)$
mixed state $\rho$, the concurrence $\mathcal{C}(\rho)$ satisfies
\begin{equation}  \label{g2}
\sqrt{\frac{2}{m(m-1)}}(\Vert \rho ^{T_{A}}\Vert -1)\leq\mathcal{C}(\rho).
\end{equation}
In our considered qubit system, $m=n=2$. Therefore it follows from Eqs.(\ref%
{c}) and (\ref{g2}) that $\mathcal{N}\leq \mathcal{C}$,
implying the negativity is never greater than the concurrence in this case. Thus for the state $%
|\phi \rangle _{ABC}$ we have
\begin{equation}  \label{g3}
\begin{array}{ccc}
\mathcal{N}_{AB}\leq \mathcal{C}_{AB} , & \mathcal{N}_{AC}\leq \mathcal{C}%
_{AC}. &
\end{array}%
\end{equation}
Observing Eqs.(\ref{a}), (\ref{g}) and (\ref{g3}), the conclusion in Eq.(\ref%
{g1}) can be proved.

In a similar way, if one takes the different focus $B$ and $C$, the
following monogamy inequalities
\begin{equation}  \label{g11}
\mathcal{N}^2_{BA}+\mathcal{N}_{BC}^2\leq \mathcal{N}^2_{B(AC)},
\end{equation}
and
\begin{equation}  \label{g12}
\mathcal{N}^2_{CA}+\mathcal{N}_{CB}^2\leq \mathcal{N}^2_{C(AB)},
\end{equation}
hold also.

\begin{figure}[tbp]
\includegraphics[width=0.40\textwidth,height=0.20\textheight]{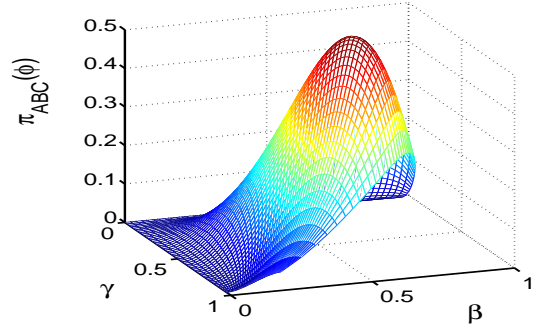}
\caption{(color online). The three-$\pi$ for the state in Eq.(9) as
a function of the coefficients $\beta$ and $\gamma$. Only two
coefficients  are independent since
$|\alpha|^2+|\beta|^2+|\gamma|^2=1$. $\pi_{ABC}(\phi)$ is always
greater than zero and reaches maximal value
$\frac{4}{9}\left(\sqrt{5}-1\right)$ when
$\alpha=\beta=\gamma=\frac{1}{\sqrt{3}}$, which, to a certain
degree, demonstrates why the $W$ state is maximally entangled.}
\end{figure}

Now one naturally would like to care about the tightness of the
monogamy inequality in Eq.(\ref{g1}). All pure three-qubit states
can be sorted into six classes through stochastic local operation
and classical
communication(SLOCC)\cite{due}. (1) $A$--$B$--$C$ class including product states; (2) $%
A$--$BC$, (3) $B$--$AC$, and (4) $C$--$AB$ classes including
bipartite entanglement states; (5) $W$ and  (6) GHZ classes
including genuine tripartite entanglement
states. For the first four classes it is easy to verify that Eqs.(\ref{g1}), (%
\ref{g11}), and (\ref{g12}) turn out to be an equality, being the same to the
CKW inequality. However, it is different for the $W$ class. For the
following pure state of $ABC$:
\begin{equation}  \label{h}
|\phi\rangle=\alpha|100\rangle+\beta|010\rangle+\gamma|001\rangle,
\end{equation}
which belongs to the $W$ class. Substituting $\mathcal{N}^2_{AB}=4\alpha^2%
\beta^2+2\gamma^4-2\gamma^2\sqrt{\gamma^4+4\alpha^2%
\beta^2}$, $\mathcal{N}^2_{AC}=4\alpha^2\gamma^2+2\beta^4-2\beta^2\sqrt{\beta^4+4\alpha^2\gamma^2}$, and $\mathcal{N}%
^2_{A(BC)}=4\alpha^2(\beta^2+\gamma^2)$ into Eq.(\ref{g1}) we have $%
\gamma^4+\beta^4< \gamma^2\sqrt{\gamma^4+4\alpha^2\beta^2}+\beta^2\sqrt{%
\beta^4+4\alpha^2\gamma^2}$, resulting in that the inequality in
Eq.(\ref{g1}) is strict due to $\alpha\neq 0$, $\beta\neq 0$, and
$\gamma\neq 0$, while the CKW inequality can only be an equality for
the $W$ class\cite{due}.

Having seen that both the equality and inequality in Eq.(\ref{g1})
can be satisfied by some three-qubit states, we can define the
residual entanglement, which is referred to as the three-$\pi$ in
order to distinguish from the three-tangle in the following main
results of this paper.

\section{\ Three-$\pi$ entanglement measure}
 The difference between the two sides of Eq.(\ref{g1}%
) can be interpreted as the residual entanglement
\begin{equation}  \label{h1}
\pi_{A}= \mathcal{N}^2_{A(BC)}-\mathcal{N}^2_{AB}-\mathcal{N}_{AC}^2.
\end{equation}
Likewise, Eqs.(\ref{g11}) and (\ref{g12}) gives birth to the corresponding
residual entanglement as
\begin{equation}  \label{h2}
\pi_{B}= \mathcal{N}^2_{B(AC)}-\mathcal{N}^2_{BA}-\mathcal{N}_{BC}^2,
\end{equation}
and
\begin{equation}  \label{h3}
\pi_{C}= \mathcal{N}^2_{C(AB)}-\mathcal{N}^2_{CA}-\mathcal{N}_{CB}^2,
\end{equation}
respectively. The subscripts $A$, $B$, and $C$ in $\pi_{A}$, $\pi_{B}$, and $%
\pi_{C}$ mean that qubit $A$, qubit $B$, and qubit $C$ are taken as the
focus, respectively. Unlike the three-tangle, in general $\pi_{A}\neq
\pi_{B}\neq\pi_{C}$, which can be easily confirmed after calculating them
for the state in Eq.(\ref{h}). This indicates that the residual entanglement
corresponding to the different focus is variant under permutations of the
qubits. We take $\pi_{ABC}$ (referred to as the three-$\pi$) as the average
of $\pi_{A}$, $\pi_{B}$, and $\pi_{C}$, i.e.,
\begin{equation}  \label{j}
\pi_{ABC}=\frac{1}{3}(\pi_{A}+\pi_{B}+\pi_{C}),
\end{equation}
which thus becomes invariant under permutations of the qubits since, for
example, permutation of qubit $A$ and qubit $B$ accordingly only leads to
exchanging $\pi_{A}$ and $\pi_{B}$ with each other in $\pi_{ABC}$.

As we will prove here, the three-$\pi$ in Eq.(%
\ref{j}) is a natural entanglement measure satisfying three
necessary conditions\cite{ve}. The first condition is that the
three-$\pi$ should be local unitary (LU) invariant. After LU
transformations $U_A$, $U_B$, and $U_C$ acted separately on a pure
three-qubit state $\rho_{ABC}$, the state can read $%
\rho^{\prime}_{ABC}=U_A\otimes U_B\otimes U_C\rho_{ABC} U^{\dag}_A\otimes
U^{\dag}_B\otimes U^{\dag}_C$. It is necessary to prove that the six squared
negativities in Eq.(\ref{j}) are invariant under the three simultaneous LU
transformations. Since $\rho^{\prime}_{A}=\mathrm{Tr}_{BC}\rho^{%
\prime}_{ABC}=U_A\rho_{A}U_{A}^{\dag}$ and $\mathcal{N^{\prime}}_{A(BC)}=%
\mathcal{C^{\prime}}_{A(BC)}=\sqrt{2(1-\mathrm{Tr\rho'^2_{A}})}=%
\mathcal{N}_{A(BC)}$ , $\mathcal{N}_{A(BC)}$ is LU invariant. Similarly $%
\mathcal{N}_{B(AC)}$ and $\mathcal{N}_{C(AB)}$ are also LU invariant.
While $\rho^{\prime}_{AB}=\mathrm{Tr}_{C}\rho^{%
\prime}_{ABC}=U_A\otimes U_B\rho_{AB} U^{\dag}_A\otimes U^{\dag}_B$,
together with the property that the negativity itself is LU invariant\cite%
{pere,vidal}, leads to $\mathcal{N}(\rho^{\prime}_{AB})=\mathcal{N}%
(\rho_{AB})$. Thus $\mathcal{N}(\rho_{AB})$ is LU invariant, so are $%
\mathcal{N}(\rho_{BC})$ and $\mathcal{N}(\rho_{AC})$. Now we finish proving the first condition.

Observation of Eqs.(\ref{g1}), (\ref{g11}) and (\ref{g12}) shows that $%
\pi_{ABC}\geq 0$, thus the second condition is satisfied. Moreover, it is
easy to verify that $\pi_{ABC}= 0$ for product pure states. $\pi_{ABC}$ is
invariant under permutations of the qubits allows us to use proof outline%
\cite{due} to confirm the third condition. Let us consider local
positive operator valued measure (POVM's) for one qubit only. Let
$A_1$ and $A_2$ be two POVM elements such that
$A^{\dag}_1A_1+A^{\dag}_2A_2=I$. We can write $A_i=U_iD_iV$, with
$U_i$ and $V$ being unitary matrices, and $D_i$ being diagonal
matrices with entries $(a, b)$ and $(\sqrt{1-a^2}, \sqrt{1-b^2})$,
respectively. Consider an arbitrary initial state $|\psi\rangle$ of
qubit $A$, $B$, and $C$
with $\pi_{ABC}(\psi)$. After the POVM, $|\phi^{\prime}\rangle=A_i|\psi%
\rangle$. Normalizing them gives $|\phi_i\rangle=|\phi^{\prime}_i\rangle/%
\sqrt{p_i}$ with $p_i=\langle\phi^{\prime}_i|\phi^{\prime}_i\rangle$ and $%
p_1+ p_2=1$. Therefore $\langle\pi_{ABC}\rangle=p_1\pi_{ABC}(\phi_1)+p_2%
\pi_{ABC}(\phi_2)$. Taking into account both the fact that $%
\pi_{ABC}(U_iD_iV\psi)=\pi_{ABC}(D_iV\psi)$ due to its LU invariance
and the key fact that three-$\pi$ is also a quartic function of its
coefficients in the standard basis which can be seen from
the calculation for the state of Eq.(\ref{h}), we have $\pi_{ABC}(\phi_1)=%
\frac{a^2b^2}{p_1^2}\pi_{ABC}(\psi)$ and $\pi_{ABC}(\phi_2)=\frac{%
(1-a^2)^2(1-b^2)^2}{p_2^2}\pi_{ABC}(\psi)$. After simple algebra
calculations, we obtain $\langle\pi_{ABC}\rangle \leq
\pi_{ABC}(\psi)$, thus the third condition that the three-$\pi$
should be an entanglement monotone is satisfied.

\begin{figure}[tbp]
\includegraphics[width=0.40\textwidth,height=0.18\textheight]{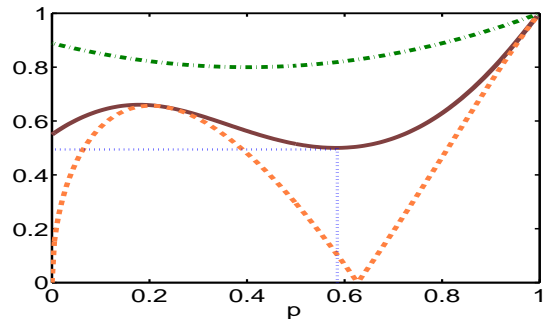}
\caption{(color online). Plot of $\pi^{(-)}_{ABC}$(solid line),
$\tau^{(-)}_{ABC}$(dashed line), and squared
$\mathcal{N}^{(-)}_{A(BC)}$ (dashed-dotted line) for the state
$|\Psi^{(-)}\rangle_{ABC}$ in Eq.(15) as a function of $p$. The
dotted line indicates the minimal value of $\pi^{(-)}_{ABC}$. The
two measures match at $p=0.2$ and $p=1$.}
\end{figure}

\section{demonstration of two examples and extension to pure multiqubit states}

In order to explicitly see the difference between the three-$\pi$
and the three-tangle we present the following two examples.

\emph{Example 1: The different classes by SLOCC.}  For the state in Eq.(\ref%
{h}) belonging to the $W$ class we get
\begin{eqnarray}
\pi _{ABC}(\phi) &=&\frac{4}{3}(\alpha ^{2}\sqrt{\alpha ^{4}+4\beta ^{2}\gamma ^{2}%
}+\beta ^{2}\sqrt{\beta ^{4}+4\alpha ^{2}\gamma ^{2}}  \notag \\
&&+\gamma ^{2}\sqrt{\gamma ^{4}+4\alpha ^{2}\beta ^{2}}-\alpha ^{4}-\beta
^{4}-\gamma ^{4})  \notag \\
&>& \tau_{ABC}(\phi)=0 \label{q}.
\end{eqnarray}
We also have performed extensive numerical calculation on
three-$\pi$ of the other states in the $W$ class and found that it
is always greater than zero (i.e., $\pi_{ABC}(W)>0$) as shown in
Eq.(\ref{q}) for the sate $|\phi\rangle$ (see also Fig.1), implying
these states have three-way entanglement also. Taking into account
that $\tau_{ABC}(W)=0$ a conclusion that the three-tangle
underestimates three-way entanglement can be drawn. For the GHZ
class we have the property that
$\pi_{ABC}(\mathrm{GHZ})=\tau_{ABC}(\mathrm{GHZ})>0$. While
$\pi_{ABC}(\mathrm{\Phi})=\tau_{ABC}(\mathrm{\Phi})=0$ for the
states $|\Phi\rangle_{ABC}$ belonging to the classes excluding the
$W$ and GHZ classes .
\begin{figure}[tbp]
\includegraphics[width=0.40\textwidth,height=0.18\textheight]{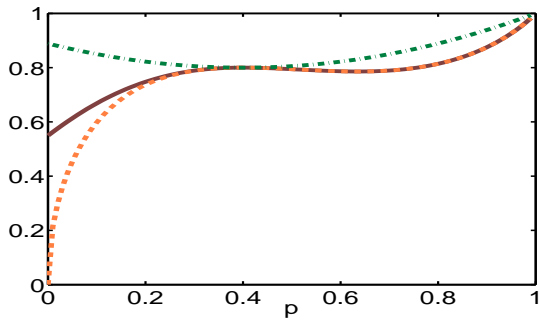}
\caption{(color online). Plot of $\pi^{(+)}_{ABC}$(solid line),
$\tau^{(+)}_{ABC}$(dashed line), and squared $\mathcal{N}^{(+)}_{A(BC)}$
(dashed-dotted line) for the state $|\Psi^{(+)}\rangle_{ABC}$ in Eq.(15) as
a function of $p$. The two measures match at $p\in [0.4,1]$,
together with squared $\mathcal{N}^{(+)}_{A(BC)}$, they match at $p=0.4$.}
\end{figure}

\emph{Example 2: Superpositions of GHZ and $W$ states.} Quantifying
of multipartite mixed states is also a fundamental issue in quantum
information theory. An optimal decompositions for the three-tangle
of mixed three-qubit states composed of a GHZ state and a $W$ state
is obtained\cite{lo}. In order to further explore the relationship
between three-$\pi$ and three-tangle we first write down superposed
state of GHZ and $W$ states
\begin{equation}\label{l}
|\Psi^{(\pm)}\rangle_{ABC}=\sqrt{p}|\mathrm{GHZ}\rangle\pm\sqrt{1-p}|W\rangle.
\end{equation}
The three-tangle for $|\Psi^{(\pm)}\rangle_{ABC}$ is known as  $\tau^{(\pm)}_{ABC}=|p^2\pm\frac{8\sqrt{6}}{9}\sqrt{p(1-p)^3}|$
 and with Eqs.(\ref{h1}-\ref{j}) we plot $\pi^{(\pm)}_{ABC}$ (see Fig.2 and 3). The two measures shows similar trend and the
fact that $\pi_{ABC}\geq \tau_{ABC}$ is shown. Notice that the
similar result was obtained also in\cite{sh}, however, their defined
residual entanglement $E=
\mathcal{N}_{A(BC)}-\mathcal{N}_{AB}-\mathcal{N}_{AC}$ is not an
entanglement measure\cite{la}. On the other hand, for the state
$|\Psi^{(-)}\rangle_{ABC}$ the location of $p$ of the minimal value
of the two measures does not match(see Fig.2), i.e., when
$p\approx0.58$ the extremely minimal $\pi^{(-)}_{ABC}\approx0.5$
which is smaller than
$\pi_{ABC}(W)=\frac{4}{9}\left(\sqrt{5}-1\right)\approx0.55$ being
equal to $\pi^{(-)}_{ABC}$ when $p=0$. But $\tau^{(-)}_{ABC}=0$ when
$p=\frac{4\sqrt[3]{2}}{3+\sqrt[3]{2}}\approx0.63$ for the state
$|\Psi^{(-)}\rangle_{ABC}$, which provides a basis on the optimal
decomposition of mixtures of the GHZ and $W$ states\cite{lo}. In a
similar way, we can also achieve optimal decomposition of such mixed
states for the three-$\pi$\cite{ou}. Note that for mixed three-qubit
states of $ABC$, the monogamy inequality Eq.(\ref{g1}) turns out to
be
\begin{equation}  \label{o}
\mathcal{N}^2_{AB}+\mathcal{N}_{AC}^2\leq \mathrm{min}[\mathcal{N}^2_{A(BC)}],
\end{equation}
which has to be minimized for all possible decomposition of
$\rho_{ABC}$. The other inequalities in Eqs.(\ref{g11}) and
(\ref{g12}) need the same manipulations.

\emph{Extension to pure multiqubit states.} The general CKW
inequality for the case of $n$ qubits is proved\cite{osborne}. Our
monogamy inequality  can also do this. Denote $n$ qubits by $A_1,
A_2, ...A_n$. Eq.(\ref{g}) may generalize to
$\mathcal{N}_{A_1(A_2A_3...A_n)}=\mathcal{C}_{A_1(A_2A_3...A_n)}$.
Considering the fact that $\mathcal{N}\leq\mathcal{C}$ for mixed
two-qubit states and the general CKW inequality\cite{osborne}, we
prove that
\begin{equation}\label{p}
\mathcal{N}^2_{A_1A_2}+\mathcal{N}^2_{A_1A_3}+\cdot\cdot\cdot+\mathcal{N}^2_{A_1A_n}
\leq \mathcal{N}^2_{A_1(A_2A_3...A_n)},
\end{equation}
which may also be used to study the entanglement for a wide class of
complex quantum systems\cite{osborne}. The general monogamy
inequality corresponding to the different focus has a similar form
in Eq.(\ref{p}).

\section{Conclusions}

 Summarizing, we proved a monogamy inequality in terms of negativity
 such that the three-$\pi$ is defined so as to quantify the residual entanglement for three-qubit states.
 The three-$\pi$ is shown to be a natural entanglement measure and
 can be extended to mixed states and general pure $n$-qubit states. The three-way entanglement for the $W$ and GHZ
 classes quantified by the three-$\pi$ always exists, while
 the three-tangle is zero for the $W$ class. Compared to the three-$\pi$, the three-tangle generally underestimates the
 entanglement. Note that the monogamy inequality for distributed Gaussian entanglement in terms of negativity was also
  established\cite{adesso} and the information-theoretic measure of genuine multiqubit entanglement based on bipartite
 partitions was introduced\cite{zhou}. Therefore, further investigation by using the results in this paper will help us deeply understand genuine multipartite entanglement.

\section*{Acknowledgement}

This work was supported in part by innovative grant of CAS.  The
author Y.C.O. was supported from China Postdoctoral Science
Foundation and the author H.F. was also partly supported by 'Bairen'
program and NSFC grant.


\begin{references}
\bibitem{la} For example, when $\alpha^2=0.9$ and $\beta^2=\gamma^2=0.05$ for the state in Eq.(\ref{h}),
we get $\mathcal{N}_{AB}+\mathcal{N}_{AC}\approx
0.75>0.6=\mathcal{N}_{A(BC)}$, leading to $E<0$.
\end{references}
\end{document}